\documentclass[epj]{svjour}
\usepackage{graphicx}
\graphicspath{ {./images/} }
\usepackage{amsmath}
\usepackage{appendix}
\begin{document}
\title{Deformation quantization in FLRW geometries}

\author{Alfonso F. Bobadilla \inst{1}\thanks{alfofe08@ucm.es} \and Jose A.R. Cembranos\inst{1,2}
\thanks{cembra@ucm.es}
}                     
%
%
\institute{Departamento de Física Teórica, Universidad Complutense de Madrid, Plaza de Ciencias 1, Facultad de Ciencias Físicas, 28040, Madrid, Spain. \and Institute of Particle and Cosmos Physics (IPARCOS), Universidad Complutense de Madrid, Plaza de Ciencias 1, Facultad de Ciencias Físicas, 28040, Madrid, Spain.}
%
%
\abstract{
We investigate the application of deformation quantization to the system of a free particle evolving within a universe described by a Friedmann-Lemaître-Robertson-Walker (FLRW) geometry. This approach allows us to analyze the dynamics of classical and quantum phase-space distributions in curved spacetime. We demonstrate that when the curvature of the spatial sections is non-zero, the classical Liouville equation and its quantum counterpart, represented by the Moyal equation, exhibit distinct behaviors. Specifically, we derive a semi-classical dynamical equation that incorporates curvature effects and analyze the evolution of the Wigner quasi-distribution function in this cosmological context. By employing a perturbative approach, we elaborate on the case of a particle described by a spherically symmetric Wigner distribution and explore the implications for phase-space dynamics in expanding universes. Our findings provide new insights into the interplay between quantum mechanics, phase-space formulations, and cosmological expansion, highlighting the importance of deformation quantization techniques for understanding quantum systems in curved spacetime.
} 
%


\maketitle

\section{Introduction}

Among the various formulations of classical mechanics, \textit{Hamiltonian mechanics} holds a prominent place due to its elegant mathematical structure, which lends itself naturally to the process of quantization. This formalism is built upon the framework of \textit{symplectic geometry}, where a symplectic manifold $(\mathcal{M}, \omega)$ consists of a differentiable manifold $\mathcal{M}$ equipped with a closed, non-degenerate 2-form $\omega$. This symplectic form enables the construction of a Poisson algebra of smooth functions on $\mathcal{M}$ and facilitates the description of system dynamics through Hamilton's equations. The evolution of a mechanical system can thus be seen as the integral curves of the Hamiltonian vector field $X_H$ associated with the Hamiltonian function $H$.

Indeed, the symplectic form allows us to establish a map from smooth functions to vector fields on $\mathcal{M}$, \cite{arnol2013mathematical,souriau1997structure}. Under this correspondence, the function $f \in \mathcal{C}^{\infty} (\mathcal{M})$ maps to the unique field $X_f$ that satisfies $\iota_{X_f} \omega = df$. With this map, naturally provided by the symplectic form, one can construct the following binary operation on $\mathcal{C}^{\infty} (\mathcal{M})$: $f,g \mapsto \{f,g\} := X_f(g)$ \cite{souriau1997structure}. The bracket `` $\{ \ , \ \}$ '' together with the usual product of functions `` $ \cdot$ '', endow the set of smooth functions with the structure of a Poisson algebra, $(\mathcal{C}^{\infty} (\mathcal{M}),\{ \, , \, \},\cdot)$, that is, an associative algebra with respect to `` $ \cdot$ '' and a Lie algebra with respect to the bracket `` $\{ \ , \ \}$ ''; both operations satisfying Leibniz identity, \cite{bates1997lectures}. Now we turn our attention to the special case of cotangent bundles. This class of manifolds naturally carry a special one-form $\theta$, the canonical one-form, whose exterior derivative, $\omega := -d\theta$, is closed $(d \omega = -d^2 \theta = 0)$ and non-degenerate and, thus, is a valid symplectic form on the cotangent bundle \cite{frankel2011geometry}.

\

The Hamiltonian formulation of mechanics is built completely on top of this result. If one interprets the manifold $\mathcal{Q}$ as the configuration space of a given mechanical system, that is, the set of all possible positions $(q^i)$, the cotangent bundle $T^*\mathcal{Q}$ can be seen as the set of all possible positions and momenta $(q^i,p_i)$, the phase space of the system. Observables are represented by smooth functions of the Poisson algebra $\mathcal{C}^\infty(T^*\mathcal{Q})$. The Hamiltonian formalism then exploits the symplectic structure of this phase space to describe the evolution of the system: trough the correspondence between functions and vector fields, one identifies the possible physical trajectories of the system with the integral curves of the field $X_H$ associated to a certain function $H$, the \textit{Hamiltonian} of the system, \cite{arnol2013mathematical,souriau1997structure,frankel2011geometry,johns2011analytical}.

In Hamiltonian mechanics, one can choose to describe the state of the system in several ways. The most immediate one is to describe the system by identifying its physical state with a point $(q^i,p_i)$ of its phase space, which evolves along a solution curve $(q^i(t),p_i(t))$ of Hamilton's equations \cite{arnol2013mathematical,souriau1997structure,frankel2011geometry}.
Nonetheless, one can exploit the symplectic structure of phase space to construct an alternative formalism. Instead of using phase space points, $(q^i,p_i)$, one can describe the state of the system with probability distribution over phase space, the so called Liouville function, $\rho$. The Liouville function characterizes the system in the sense that its integral over some region of phase space gives the probability of finding the system in it ($\rho$ must be normalized to 1) \cite{landau2013course,greiner2012thermodynamics} and, in its time development, satisfies the Liouville equation,
\begin{equation}\label{EqLiouville}
    \frac{\partial \rho}{\partial t} + \{\rho,H\} = 0.
\end{equation}
The density function formalism is somewhat more general than the point-based one reviewed above, since it allows for the description of ensembles of multiple systems, or systems with non perfectly defined initial conditions. 

This result is especially interesting when pursuing a quantization process. Since a quantum system presents an inherent indeterminacy on the values of its spacial coordinates and momenta, while the point-based description results inadmissible, the density function formalism seems to be more adequate to the circumstance. It is through the alteration of this last formalism that the phase space formulation of quantum mechanics is constructed \cite{groenewold1946principles,moyal1949quantum,curtright2013concise}. Recent developments have further enriched this approach, such as the application of the Wigner formalism to the Aharonov–Bohm effect \cite{Cembranos2024}, the use of the Segal–Bargmann formalism to study Wehrl entropy in entangled systems \cite{AlonsoLopez2023},  the formulation of functional quantum field theory in phase space \cite{Cembranos2021}, and the Wigner formalism on black hole geometries \cite{Garcia:2024inq}. These developments demonstrate how versatile phase-space methods are in describing quantum phenomena, effectively addressing a wide range of systems, even those within curved geometries.

\section{Deformation quantization in curved phase spaces}
The most remarkable feature of a quantum theory is the non-commutativity of its observables. The approach taken by Deformation Quantization (DQ) is to \textit{deform} the operations $\cdot$ and $\{ \ , \ \}$ of the algebra of classical observables of a given system to get a non-commutative algebra that represents the algebra of observables of the quantum theory of the system \cite{curtright2013concise}. We then say that we have deformed the original manifold into a non-commutative phase space. The concept of interpreting quantum mechanics as a deformation of classical mechanics originates from the work of Flato, Lichnerowicz, and their collaborators, who pioneered this approach and developed the associated techniques in the late 1970s \cite{flato1976deformations,bayen1977quantum}.


If $\mathcal{M}$ is a symplectic manifold, a deformation of its Poisson algebra $(\mathcal{C}^{\infty}(\mathcal{M}) \, , \ \cdot \ , \, \{ \ , \ \} \, )$ is a family of algebras $(\mathcal{C}^{\infty}(\mathcal{M}) \, , \star , \{\!\{ \ , \ \}\!\})$, parameterized by a variable $\hbar$, where the binary operation $\star$, known as star product, is the deformation of the associative product $\cdot$ of the original algebra, and the bracket $\{\!\{ \ , \ \}\!\}$ is the deformation of the Poisson bracket $\{ \ , \ \}$ of the original algebra \cite{bates1997lectures}. We require that when $\hbar = 0$ we recover the original Poisson algebra. That is, for $f$ and $g$ in $\mathcal{C}^{\infty}(\mathcal{M})$, if we write $f \star g$  and $\{\!\{ f , g \}\!\}$ as a power series on the formal parameter $\hbar$,

\

\ \textit{i})   $f \star g = \sum_{n=0}^{\infty} c_n(f,g)\hbar^n$, with $c_0(f,g) = f \cdot g$\,.

\

\ \textit{ii})   $\{\!\{f,g\}\!\} = \frac{1}{i\hbar}(f \star g - g \star f)$\,,

\

where $c_i( \ , \ )$ are differential operators on its two arguments. The star product $\star$ must also verify the condition

\

\ \textit{iii})   $c_1(f,g)-c_1(g,f) = i \, \{f,g\}$\,.

\

Fedosov and Kontsevich showed that deformation quantization (DQ) could be generalized to arbitrary symplectic and Poisson manifolds, as long as an appropriate connection is defined \cite{kontsevich2003deformation,fedosov1996deformation}. This extension is particularly useful for studying mechanical systems in curved spacetimes, where the phase space naturally acquires curvature and deviates from the conventional flat phase-space structure.

If we are to reproduce the quantum commutation relations between the observables of the system, only a special class of deformations of the Poisson algebra of classical observables will work. Among them, there is a star product whose construction is particularly simple: the Moyal product. It is defined as \cite{tillman2008fedosov,hirshfeld2002deformation}
\begin{equation} \label{ProdMoyalDQ}
    \star := \exp(i \hbar \Pi / 2) = 1 + \frac{i\hbar}{2} \Pi + \frac{1}{2}\left(\frac{i\hbar}{2}\right)^2\Pi ^2 + ... \ ,
\end{equation}
with $\Pi = \overleftarrow{D}_\alpha\omega^{\alpha \beta}\overrightarrow{D}_\beta$, where $D$ represents a covariant derivative defined using a proper phase-space connection. Throughout the article, Greek indices ($\alpha$, $\beta$, ...) will be used to represent phase-space coordinates in a uniform manner. That is $\xi^\alpha = q^\alpha$ if $\alpha \leq n$ and $\xi^\alpha = p_{\alpha-n}$ if $\alpha > n$.  The successive powers of $\Pi$ are defined as
\begin{equation}
    \Pi^{\, k} := \overleftarrow{D}^{(k)}_{\alpha_1 ... \alpha_k}\omega^{\alpha_1\beta_1} ... \ \omega^{\alpha_k\beta_k}\overrightarrow{D}^{(k)}_{\beta_1 ... \beta_k},   
\end{equation}
where $D^{(k)}_{\alpha_1 ... \alpha_k}f$ is defined as the component ($\alpha_1, \, ... \, , \, \alpha_k$) of the the action of the $k$-fold composition of the covariant derivative with itself on $f$: $[(\,D \circ ... \circ D\,)f \, ]_{\alpha_1 ... \alpha_k}$. Since it is a tensor quantity, its definition remains covariant under general coordinate transformations in phase space. Additionally, if we introduce the definition \cite{tillman2008fedosov}
\begin{equation}
    \{\!\{f,g\}\!\} := (f \star g - g \star f)/i\hbar,
\end{equation}
it can be seen that both operations constitute an admissible deformation of $\mathcal{C}^\infty(\mathcal{M})$. Most importantly, by construction, this deformation reproduces the adequate quantum commutation relations between position and momenta,
\begin{equation}
    q^i \star p_j - p_j \star q^i = i\hbar \delta^i_j,
\end{equation}
and the rest equal to zero. Once we have an adequate deformation of the phase space, we can smoothly shift between the classical and quantum theories of the system. The classical results are translated into the quantum theory, replacing products and brackets by their respective deformations $\star$ and $\{\!\{ \ , \ \}\!\}$. For example, with the change $\{ \ , \ \} \mapsto \{\!\{ \ , \ \}\!\}$ the classical evolution equation (\ref{EqLiouville}) transforms into the Moyal equation
\begin{equation}\label{Moyaleq}
    \frac{\partial \rho}{\partial t} + \{\!\{\rho,H\}\!\} = 0.
\end{equation}

On the other hand, the classical limit is achieved taking $\hbar \rightarrow 0$ in the equations of the quantum theory, which ensures that we recover commutativity between observables and a time evolution governed by Liouville equation.
\section{FLRW geometries}
In this section, we describe the application of DQ to the system of a non-relativistic particle on an expanding universe described by a Friedman-Lemâitre-Robertson-Walker (FLRW) geometry. In the comoving coordinate system $(t,r,\theta,\phi)$, the FLRW metric reads \cite{hawkingellis}
\begin{equation}
    ds^2 = dt^2 - a^2(t) \left( \frac{dr^2}{1-kr^2} + r^2(d\theta^2 + \cos^2 \theta \ d\phi^2) \right),
\end{equation}
being $k$ the curvature of the spatial sections and $a(t)$ the scale factor. A free particle, without any kind of internal structure, has just three degrees of freedom. For simplicity, we will work on the aforementioned comoving coordinate system, in which space and time split conveniently. Therefore, by treating the time coordinate $t$ as an external parameter, we take a spatial slice with Riemannian 3-metric that scales with $t$,
\begin{equation}
    g^{(3)}_t = a^2(t)\left( \frac{dr^2}{1-kr^2} + r^2(d\theta^2 + \cos^2 \theta \ d\phi^2) \right),
\end{equation}
as the configuration space of the free particle, $\mathcal{Q}$, with coordinates $(r,\theta,\phi)$. With this choices, the phase space of the particle is the cotangent bundle $T^*\mathcal{Q}$.

At low energies, the Hamiltonian that generates the non-relativistic evolution of the free particle is
\begin{equation} \label{Hamiltoniano}
    H = \frac{1}{2m\,a^2}\left((1-kr^2)p_r^2+ \frac{p_{\theta}^2}{r^2} + \frac{p_{\phi}^2}{r^2\cos^2\theta}\right).
\end{equation}

If we characterize the state of the particle with a probability distribution $\rho$, it must satisfy Liouville equation (\ref{EqLiouville}) for the former Hamiltonian.

We will deform this classical theory through the Moyal product. To obtain the quantum theory of the free particle, one first needs to 

\

\ 1) Specify a connection $D$ on its phase space, $T^*\mathcal{Q}$.

\

\ 2) Construct the Moyal star product as in (\ref{ProdMoyalDQ}).

\

\ 3) Obtain the new dynamics by inserting the Hamiltonian (\ref{Hamiltoniano}) into the Moyal equation.

\

Since the configuration space, $\mathcal{Q}$, is a spatial section with an induced riemannian 3-metric, it naturally carries the Levi-Civita connection $\nabla$ associated to $g^{(3)}_t$. Now remains the problem of extending a connection in the configuration space to the whole phase space. Fortunately, given a differentiable manifold $\mathcal{Q}$ with connection $\nabla$, there exist different mathematical procedures to extend it to a connection $D$ over the whole of its cotangent bundle $T^*\mathcal{Q}$ \cite{ConnectionsTM}. For our purposes, we will follow the procedure used in \cite{tillman2008fedosov} to extend the Levi-Civita connection of the configuration space to the unique phase-space connection that is torsion-free and compatible with the symplectic form ($D \otimes \omega = 0$). The extension procedure is the following:

\

1) Let $(q^i,p_i)$ be a canonical coordinate system for $T^*\mathcal{Q}$, so that the $(q^i)$ are coordinates for $\mathcal{Q}$. In the coordinates $(q^i)$, the coefficients fo the conection $\nabla$ on $\mathcal{Q}$ verify  $\nabla \otimes dq^i = - \Gamma^i_{jk} dq^k \otimes dq^j$. From the 1-form basis $\{dq^i,dp_i\} = \{d\xi^\alpha\}$ we construct the following non-holonomic 1-form basis:
$\theta^{(i)} := dq^i = d\xi^i$ and $\theta^{(i+n)} := dp_i - \Gamma^j_{ik}p_jdq^k = d\xi^{i+n} - \Gamma^j_{ik}\xi^{j+n}d\xi^k$.

\

2) The connection $D$ on the cotangent bundle is defined through its action on the non-holonomic 1-form basis $\{\theta^{(\alpha)}\}$. Let $R$ be the Riemann curvature tensor of $\nabla$. The connection $D$ such that $D \otimes \,  \theta^{(i)} := - \Gamma^i_{jk} \theta^{(k)} \otimes \, \theta^{(j)}$ and $D \otimes \, \theta^{(i+n)} := -\frac{8}{3}R^l_{(ij)k} \, p_l \, \theta^{(k)} \otimes \, \theta^{(j)} + \Gamma^j_{ik} \, \theta^{(k)} \otimes \, \theta^{(j+n)}$ turns out to be the only one that guarantees compatibility with the symplectic form $D \otimes \omega = 0$.

\

3) Once the action of $D$ on the non-holonomic basis $\{\theta^{(\alpha)}\}$ has been characterized, by a change of basis to the coordinate basis $\{dq^i,dp_i\}$, the action of $D$ on the 1-forms $\{dq^i,dp_i\}$ can be obtained.

\



Once the geometry of phase space is established, we can construct its Moyal equation following (\ref{Moyaleq}). Since we have the connection coefficients, $\Gamma^\alpha_{\beta \gamma}$, it is possible to explicitly calculate the terms of the series (\ref{ProdMoyalDQ}). For example, the action of $D^{(2)}$ on an arbitrary function $f$ is $D^{(2)}_{\alpha_1\alpha_2} \, f = [(D \circ D) \, f]_{\alpha_1\alpha_2} = \partial^2_{\alpha_1\alpha_2}f - \Gamma^\beta_{\alpha_2\alpha_1}\partial_{\beta} f$, and higher order terms can be calculated in a similar fashion, albeit with much more computational effort.

\

The explicit form of the Moyal equation for the free particle is obtained by introducing the Hamiltonian (\ref{Hamiltoniano}) in
\begin{equation}
    \frac{\partial \rho}{\partial t} + \{\!\{\rho,H\}\!\} = 0.
\end{equation}

If we expand the Moyal bracket as a power series on $\hbar$,
\begin{equation} \label{BrackMoyal series}
 \begin{split}
    & \{\!\{\rho,H\}\!\}=\\
    &\ \frac{2}{\hbar}\left[\rho\left(\frac{\hbar}{2} \overleftarrow{D}_\alpha \omega^{\alpha\beta} \overrightarrow{D}_\beta - \frac{\hbar^3}{8}\overleftarrow{D}_{\alpha_I}^{(3)}\omega^{\alpha_I\beta_J}\overrightarrow{D}_{\beta_J}^{(3)} + ...\right)H\right]= \\
    &\ \{\rho,H\} - \rho\left(\frac{\hbar^2}{4}\overleftarrow{D}_{\alpha_I}^{(3)}\omega^{\alpha_I\beta_J}\overrightarrow{D}_{\beta_J}^{(3)}\right)H + \mathcal{O}(\hbar^4),
\end{split}   
\end{equation}
where $D^{(3)}_{\alpha_I}=D^{(3)}_{\alpha_1\alpha_2\alpha_3}$ and $\omega^{\alpha_I\beta_J} = \omega^{\alpha_1\beta_1}\omega^{\alpha_2\beta_2}\omega^{\alpha_3\beta_3}$. We can then rewrite the Moyal equation as
\begin{equation} \label{EqMoyal series}
\frac{\partial \rho}{\partial t} + \{\rho,H\} - \rho\left(\frac{\hbar^2}{4}\overleftarrow{D}_{\alpha_I}^{(3)}\omega^{\alpha_I\beta_J}\overrightarrow{D}_{\beta_J}^{(3)}\right)H + \mathcal{O}(\hbar^4) = 0.
\end{equation}

An important result to note is that, when $k = 0$, the Moyal bracket reduces to the Poisson bracket. This becomes evident in the flat case, where it is always possible to use Cartesian coordinates $(x,y,z,p_x,p_y,p_z)$, causing the connection coefficients to vanish and the covariant derivatives to simplify to partial derivatives. In these coordinates, the Hamiltonian of a free particle is a polynomial of degree 2 in the momenta. The first quantum corrections involve third-order derivatives, which all vanish, leading to \(\{\!\{\rho, H\}\!\} = \{\rho, H\}\). Consequently, any differences between the classical and quantum evolution for a non-relativistic free particle can only appear in universes where \(k \neq 0\).

The classical limit of the Moyal equation appears specially evident in (\ref{EqMoyal series}), since when $\hbar \rightarrow 0$, the only contribution that survives is the zeroth order term, the Poisson bracket, and the classical Liouville equation is recovered. Nevertheless, despite the formal resemblance between the Liouville and Moyal equations, solving the latter is significantly more challenging. While the Liouville equation is a first-order partial differential equation, the Moyal equation incorporates pseudo-differential operators of potentially infinite order. To avoid dealing with the infinite series of terms in the Moyal bracket, we can apply the following approximation.

By expanding the Moyal bracket in a power series on $\hbar$, the higher order terms can be interpreted as successive quantum corrections to the Poisson bracket (zeroth-order term), which are responsible for the quantum behavior of the system. If the scale of the characteristic actions of the system is much greater than $\hbar$, one can perform the semi-classical approximation of retaining just the first quantum correction in the Moyal bracket (\ref{BrackMoyal series}), which is proportional to $\hbar^2$. Although we are renouncing to obtain a full quantum description, the inclusion of this term may allows us to glimpse the appearance of non-classical behaviors of the system. If we insert the free particle Hamiltonian (\ref{Hamiltoniano}) we obtain the semi-classical evolution equation for a free particle on a FLRW universe. The explicit form of the equation can be found in Appendix A.

\section{The Wigner function for curved FLRW geometries}
Even when working at order $\mathcal{O}(\hbar^2)$, just the first quantum correction involves derivatives of $\rho$ of the third, second and first order with respect to the six phase-space coordinates, which renders solving the dynamical equation for an arbitrary initial condition extremely challenging. In this section, we elaborate the special case of a free particle in ``free radial propagation'', that is, with its momenta $p_\theta$ and $p_\phi$ perfectly defined at 0.

As a consequence of the uncertainty principle, there is a total ignorance regarding the direction of propagation, and the solution is spherically symmetric:
\begin{equation} \label{spherical}
    \rho(r,\theta,\phi,p_r,p_\theta,p_\phi,t) = R(r,p_r,t) \, \delta(p_\theta) \, \delta(p_\phi).
\end{equation}

This considerably reduces the number of terms in the equation. We can simplify the problem further by assuming a low-curvature regime. The quantity $|k|^{-1/2}$ represents a very large distance, and, for example, in the case of our Universe, this length scale exceeds the size of the observable universe. Therefore, as long as we restrict ourselves to studying times in which the particle has not had the opportunity to travel a distance comparable to $|k|^{-1/2}$, we can disregard all terms with higher powers of $k$ and retain only linear terms. The semi-classical equation for the non-trivial part of the quasi-distribution reduces to
\begin{equation} \label{EcSemiMoyalFLRW}
    \frac{\partial R}{\partial \tau} + \frac{1}{m}\left((1-kr^2)p_r\frac{\partial R}{\partial r} + krp_r^2\frac{\partial R}{\partial p_r} + \frac{4\hbar^2}{9}\frac{kp_r}{r}\frac{\partial^2R}{\partial p_r^2}\right) = 0,
\end{equation}
where we have introduced a new time coordinate $\tau$ such that $d\tau = dt/a^2$.

\begin{figure}[h!]
  \includegraphics[scale=0.475]{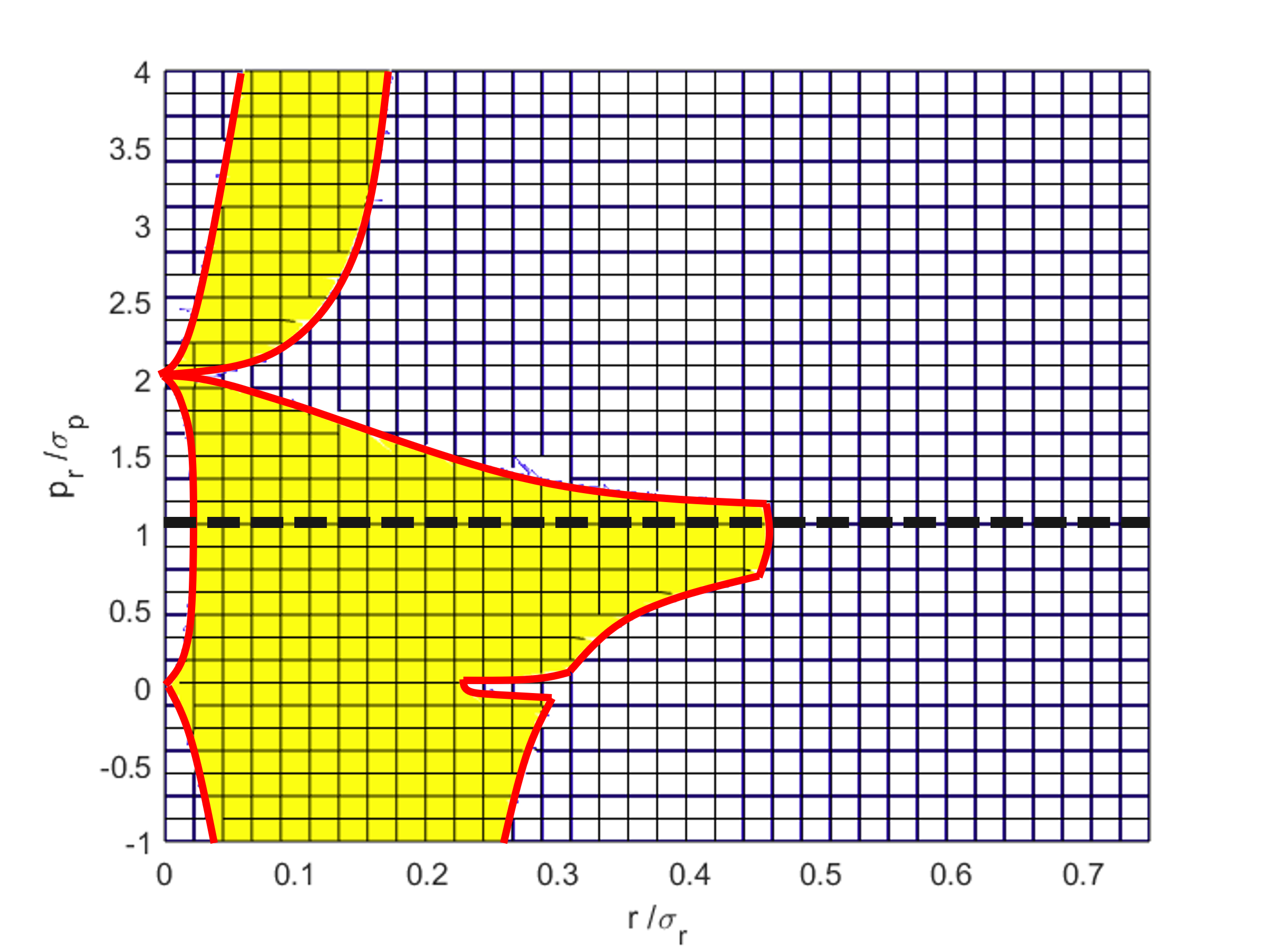}
\caption{Projection of the phase space of a free particle in a FLRW metric into the normalized radial coordinate $r$ and radial momentum $p_r$. The yellow color shows the validity region where we can rewrite the semi-classical equation as the classical one plus a quantum correction. In the case of a FLRW geometry with curvature parameter $k=\sigma_r^{-1/2} \times 10^{-4}$ and a particle with an initial Gaussian distribution with the momentum centered in $p_0 = 1 \times \sigma_p$ (the horizontal dashed line corresponds to the region $p_r = p_0$).}
\label{region}
\end{figure}

We study the particular case of a free radial propagation with an initial distribution given by a Gaussian distribution of the form 
\begin{equation}\label{IniCo}
\mathcal{R}(r,p_r) = \mathcal{N} e^{-\left(\frac{r^2}{2\sigma_r^2} + \frac{(p_r - p_0)^2}{2\sigma_{p_r}^2}\right)},
\end{equation}
where \(\sigma_r\) and \(\sigma_{p_r}\) represent the standard deviations in position \(r\) and momentum \(p_r\), respectively, and \(p_0\) is the center of the momentum distribution. The normalization constant \(\mathcal{N}\) ensures that the distribution is properly normalized.

By plugging this initial condition into (\ref{EcSemiMoyalFLRW}) we get an idea of the size and importance of the different terms of the equation
\begin{align}
    (1-kr^2)p_r\frac{\partial \mathcal{R}}{\partial r} & \ \longrightarrow \ -\frac{rp_r}{\sigma^2_r}\mathcal{R} + k\frac{r^3p_r}{\sigma^2_r}\mathcal{R},\label{firstterm}\\
    krp_r^2\frac{\partial \mathcal{R}}{\partial p_r} & \ \longrightarrow \ k\frac{rp_r^2(p_r - p_0)}{\sigma_{p_r}^2}\mathcal{R},\label{secondterm}\\\label{thirdterm}
    \frac{8\hbar^2}{3}\frac{kp_r}{r}\frac{\partial^2 \mathcal{R}}{\partial p_r^2} & \ \longrightarrow \ -\frac{4}{9}\frac{k\hbar^2}{r}\left(\frac{p_r}{\sigma^2_{p_r}} + \frac{p_r(p_r-p_0)^2}{\sigma^4_{p_r}}\right)\mathcal{R}.
\end{align}

To simplify the analysis, we focus on a region where we can neglect all terms except for the first one in (\ref{firstterm}), which represents the classical, flat evolution, and the term (\ref{thirdterm}), which includes the quantum corrections that we wish to focus on. For an initial configuration of minimal uncertainty (\(\sigma_r \sigma_p = \hbar\)), this condition is satisfied in the following region of the \((r, p_r)\) plane.

For an initial Gaussian distribution, we expect the time evolution of the solution to preserve some characteristics of the Gaussian packet behavior, at least for a short period. During the early stages of the evolution, we anticipate that the previous distinction of significant terms will remain valid in the region shown in Fig. \ref{region}. Thus, by neglecting the irrelevant terms, the semi-classical equation for a spherically symmetric solution that begins as a highly localized Gaussian distribution is, within the valid region,
\begin{equation}\label{simple}
    \frac{\partial R}{\partial \tau} + \frac{1}{m}\left(p_r\frac{\partial R}{\partial r} + \frac{4}{9}k\hbar^2\frac{p_r}{r}\frac{\partial^2 R}{\partial p_r^2}\right) = 0.
\end{equation}

If we introduce new coordinates \((\bar{\tau}, \bar{r}, \bar{p_r})\) defined by \(\bar{\tau} := \tau / m\), \(\bar{r} := r - p_r \tau / m\), and \(\bar{p_r} := p_r\), equation (\ref{simple}) transforms to
\begin{equation}\label{simpelcoord}
    \frac{\partial R}{\partial \bar{\tau}} + \Lambda\frac{\bar{p_r}}{\bar{r} + \bar{p_r}\bar{\tau}}\left(\bar{\tau}^2\frac{\partial^2R}{\partial \bar{r}^2} - 2\bar{\tau}\frac{\partial^2 R}{\partial \bar{r} \partial \bar{p_r}} + \frac{\partial^2 R}{\partial \bar{p_r}^2}\right) = 0,
\end{equation}
where $\Lambda = 4k\hbar^2/9$ groups all the constants of the equation. In the classical limit ($\Lambda \rightarrow 0$), it is possible to find a simple solution that for the initial condition \eqref{IniCo} is given by
\begin{equation}\label{simpelcoord}
   R_0(\bar{r},\bar{p}_r,\bar{\tau}) = \mathcal{N} e^{-\left(\frac{\bar{r}^2}{2\sigma_r^2} \ + \frac{(\bar{p}_r-p_0)^2}{2\sigma_{p_r}^2}\right)}\,.
\end{equation}
In the new coordinates, the Gaussian packet does not evolve with time $\bar{\tau}$, which means that the distribution spreads and advances in the original $r$ coordinate with a linear dependence on the original $\tau$. We treat the semi-classical case (small, non-zero $\Lambda$) perturbatively, by writing the solution as $R(r,p_r,\tau) = R_0(r,p_r,\tau) + \Lambda f(r,p_r,\tau)$ and neglecting terms of the order $\mathcal{O}(\Lambda^2)$ or higher. Inserting our ansatz in ($\ref{simpelcoord}$) we obtain a differential equation for the perturbation
\begin{equation}
\begin{split}
    \frac{\partial f}{\partial \bar{\tau}} =& \ -\frac{1}{\bar{r}+\bar{p_r}\bar{\tau}}\left(\bar{\tau}^2\frac{\partial^2 R_0}{\partial \bar{r}^2} - 2\bar{\tau}\frac{\partial^2 R_0}{\partial \bar{r} \partial \bar{p_r}} + \frac{\partial^2 R_0}{\partial p_r^2}\right)\\
    =& \ - \frac{1}{\bar{r}+\bar{p_r}\bar{\tau}}\bigg(\frac{\bar{\tau}^2}{\bar{\sigma}_r^2}\left(\frac{\bar{r}^2}{\sigma_r^2} - 1\right) - 2\bar{\tau}\frac{\bar{r}(\bar{p_r}-p_0)}{\sigma^2_r\sigma^2_{p_r}}\\
    &+ \ \frac{1}{\sigma_{p_r}^2}\left(\frac{(\bar{p_r} - p_0)^2}{\sigma_{p_r}^4} - 1\right)\bigg),
\end{split}
\end{equation}
whose solution, expressed in the coordinates $(r,p_r,\tau)$, is
\begin{equation}
\begin{split}\label{perturbation}
    &f(r,p_r,\tau) = - \frac{1}{2\sigma_{p_r}^3} \bigg[\bigg(\frac{(r-p_r\tau/m)^2}{\sigma_r^2}\bigg(\frac{(r-p_r\tau/m)^2}{\sigma_r^2} - 1 \\
    +& \ 2\frac{p_r(p_r-p_0)}{\sigma_{p_r}^2}\bigg) + \frac{p_r^2}{\sigma_{p_r}^2}\left(\frac{(p_r - p_0)^2}{\sigma_{p_r}^2} - 1\right)\bigg] \log\left|\frac{r}{r-p_r\tau/m}\right|\\
    -& \ \frac{p_r\tau}{2\sigma_r\sigma_{p_r}^3}\bigg[\left(\frac{(r - p_r\tau/m)^2}{\sigma_r^2} - 1\right)\frac{p_r\tau}{\sigma_{p_r}} - 2\frac{(r - p_r\tau/m)}{\sigma_r}\\
    +& \ 4\frac{p_r(p_r-p_0)(r - p_r\tau/m)}{\sigma_r\sigma_{p_r}^2}\bigg].
\end{split}
\end{equation}

\begin{figure}[h!]
  \includegraphics[scale=0.475]{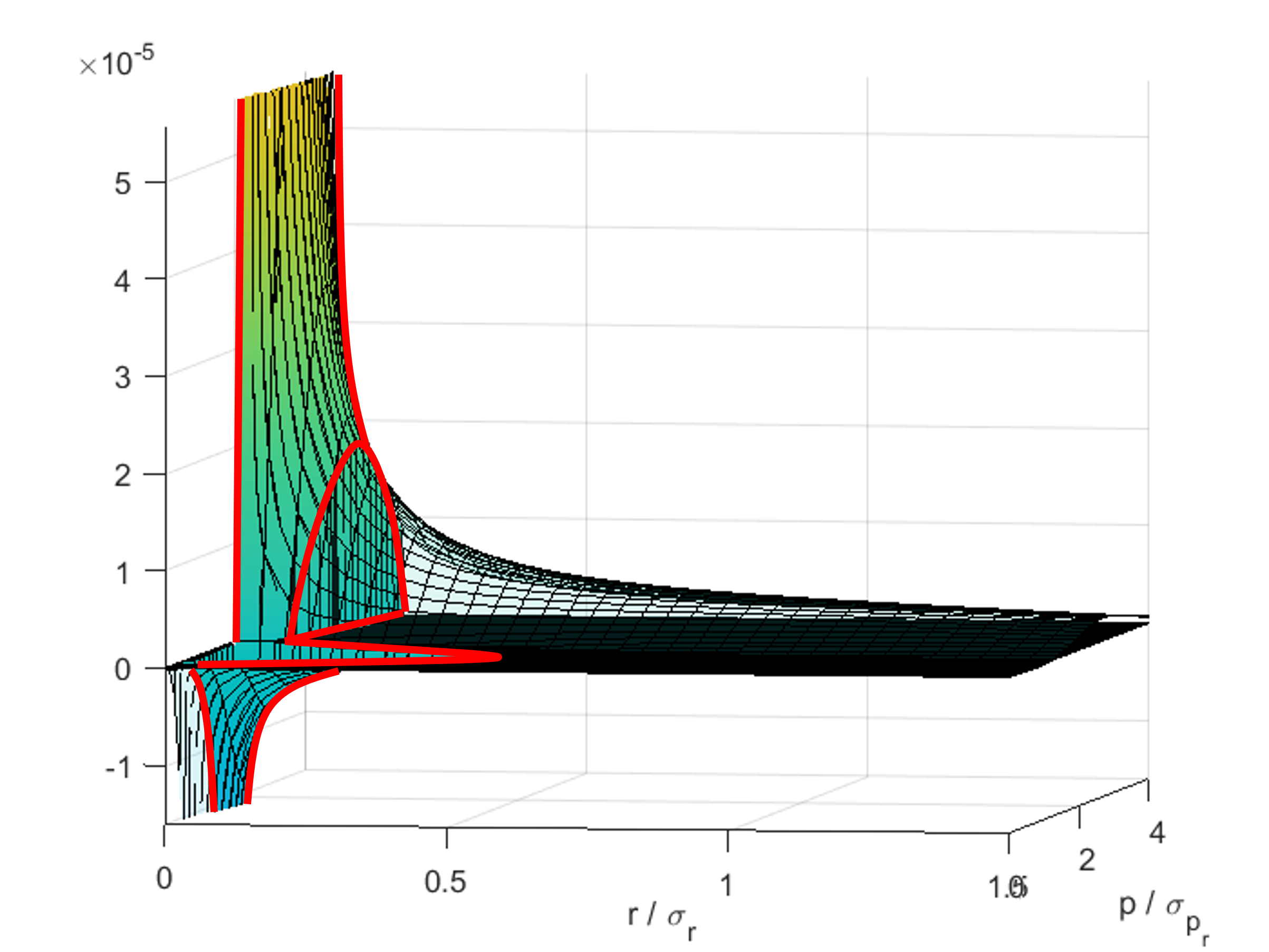}
\caption{Graphical representation of the first order quantum correction to the Wigner function  ($\Lambda f$) in the  projection of the phase space given by the $(r,p_r)$ plane, evaluated at $\tau = 0.01$.  The uncolored mesh shows $\Lambda f$ in the whole plane. The colored part (delimited by the red lines) is restricted to the region of validity shown in Fig. \ref{region}, where the discrimination of terms that permits the perturbative treatment holds.}
\label{pert}
\end{figure}

\

In Fig. \ref{pert}, it is shown the perturbative correction plotted on the region in which the discrimination of significant terms holds. The perturbative solution to the semi-classical evolution of the spherically symmetrical Gaussian packet Wigner function on a FLRW universe is then
\begin{equation}
    R(r,p_r,\tau) = R_0(r,p_r,\tau) + \frac{4}{9}k\hbar^2 \, f(r,p_r,\tau).
\end{equation}

The perturbative correction is more significant for small values of \(r\), in fact, it diverges logarithmically. For high values of \(p_r\) (\(p_r - p_0 > 2 \sigma_{p_r}\)), the correction is positive, but it is negative for shorter values (\(p_r - p_0 < 2 \sigma_{p_r}\)). This suggests that the Wigner function could reach negative values in this region, which is a clear indication of quantum effects. In any case, this change of sign lies outside the validity of the perturbative approximation, so it can only be considered as an indication.

\section{Conclusions}
In this work, we have demonstrated that DQ offers distinct conceptual and practical advantages over other quantum mechanical frameworks. One of its most compelling features is that it operates entirely within phase space, eliminating the need for additional structures like Hilbert spaces. This intrinsic phase-space formulation allows for a seamless integration of classical and quantum descriptions, facilitated by the star product, which can be expanded as a power series in the Planck constant, $\hbar$. As a result, DQ naturally supports semiclassical approximations, making it particularly well-suited for analyzing systems where quantum and classical effects coexist.

Another significant advantage of DQ is its geometric foundation, which ensures manifest covariance under general coordinate transformations in phase space. This geometric nature is particularly beneficial when dealing with systems in curved backgrounds, as it permits a consistent quantization process as long as an appropriate phase-space connection is defined. In contrast to other formalisms that may struggle with non-planar geometries, DQ accommodates these scenarios effectively, albeit with increased computational complexity. This flexibility makes DQ an invaluable tool for exploring quantum phenomena in contexts where spacetime curvature cannot be neglected.

We applied these advantages to study the quantum dynamics of a free particle in a universe described by a FLRW geometry. Our analysis revealed that in flat FLRW geometries (where the curvature parameter $k=0$), the classical and quantum dynamical equations coincide, reflecting the intuitive expectation that flat backgrounds impose no additional geometric constraints. However, when the spatial curvature is non-zero ($k \neq 0$), new terms emerge in the dynamical equation due to the contributions of the non-flat phase-space connection. These curvature-induced corrections illustrate the sensitivity of phase-space dynamics to the underlying geometry and highlight the necessity of considering geometric effects in quantum cosmology.

To better understand these corrections, we derived a semi-classical dynamical equation by expanding the full quantum equation in powers of $\hbar$ and neglecting higher-order terms. This semi-classical equation captures the essential features of the system's evolution in a curved FLRW background. In particular, we presented a perturbative solution for the case of a particle described by a spherically symmetric Wigner quasi-distribution, with the canonical momenta $p_\theta$ and $p_\phi$ set to zero initially. This highly symmetric case served as a tractable example to illustrate the formalism, though first-order perturbation theory was insufficient to detect regions where the Wigner function could become negative—a hallmark of non-classical behavior.

While the conceptual elegance of DQ is clear, the practical implementation often involves cumbersome calculations. This computational difficulty was evident throughout the latter sections of this work, especially when dealing with curved geometries where traditional simplifications, such as Bopp shifts, are not applicable. In non-planar backgrounds, the evaluation of star products becomes intricate, limiting the feasibility of exact solutions to highly symmetric systems or those amenable to approximations in specific regimes. This limitation underscores the need for developing more efficient computational techniques to handle DQ in complex geometries.

Moreover, our analysis was restricted to non-relativistic systems with a finite number of degrees of freedom. Extending DQ to relativistic settings and field theories remains a challenging but promising direction. Notable progress has been made in this area, as demonstrated by works such as \cite{dito1997deformation,zachos1999phase}, which explore the quantization of relativistic fields on infinite-dimensional symplectic manifolds. These efforts connect DQ with Algebraic Quantum Field Theory (AQFT), providing a robust framework for understanding quantum fields in curved spacetime.

In conclusion, while deformation quantization presents certain computational challenges, its conceptual clarity, geometric consistency, and adaptability to curved backgrounds make it a powerful tool for exploring quantum dynamics in both classical and cosmological contexts. Future work aimed at refining computational techniques and extending the formalism to relativistic and field-theoretic settings will further unlock the potential of DQ in addressing fundamental questions in quantum physics and cosmology.
\section*{Acknowledgments}

This work is partially supported by the project PID2022-139841NB-I00 funded by MICIU/AEI/10.13039/50110001 1033 and by ERDF/EU, and by COST
(European Cooperation in Science and Technology) Actions CA21 106, CA21136, CA22113 and CA23130. 
This work was made possible by Institut Pascal at Université Paris-Saclay with the support of the program “Investissements d’avenir” ANR-11-IDEX-0003-01, the P2I axis of the Graduate School of Physics of Université Paris-Saclay, as well as IJCLab, CEA, IAS, OSUPS, and APPEC. 

\onecolumn

\appendix

\section*{Appendix A: Semiclassical equation and low curvature regime}
\label{AppA}

The truncation at $\mathcal{O}(\hbar^2)$ of Moyal equation is $\frac{\partial \rho}{\partial t} + \{\rho,H\} - \frac{\hbar^2}{4}\left( \rho \overleftarrow{D}_{\alpha_1\alpha_2\alpha_3}^{(3)}\omega^{\alpha_1\beta_1}\omega^{\alpha_2\beta_2}\omega^{\alpha_3\beta_3}\overrightarrow{D}_{\beta_1\beta_2\beta_3}^{(3)} H \right) = 0 \,$. When $D$ is the phase-space connection of a FLRW background and $H$ is the one given in (\ref{Hamiltoniano}), then the $\mathcal{O}(\hbar^2)$ term is
\begin{equation}
\begin{split}
&4\hbar^2\left( \frac{k\,{p_\theta}^2}{3\,{r}^3\,\left(k\,{r}^2-1\right)}\frac{\partial  \rho^3}{\partial {p_r}^3}+k\,{p_\phi}^2\,\cos \theta\,\frac{\partial  \rho^3}{\partial {p_r}^3}+k^2\,p_r\,p_\theta\,{r}^4\,\cos^3\theta\,\frac{\partial }{\partial p_\theta} \frac{\partial  \rho^2}{\partial {p_r}^2}-2\,k\,p_r\,p_\theta\,{r}^2\,\cos^3\theta\,\frac{\partial }{\partial p_\theta} \frac{\partial  \rho^2}{\partial {p_r}^2}\right.\\
& -\sin \theta\,k\,{p_\phi}^2\,r\,\frac{\partial }{\partial p_\theta} \frac{\partial  \rho^2}{\partial {p_r}^2}+k^2\,p_r\,p_\phi\,{r}^4\,\cos^3\theta\,\frac{\partial }{\partial p_\phi} \frac{\partial  \rho^2}{\partial {p_r}^2} -2\,k\,p_r\,p_\phi\,{r}^2\,\cos^3\theta\,\frac{\partial }{\partial p_\phi} \frac{\partial  \rho^2}{\partial {p_r}^2}-k^2\,{p_r}^2\,{r}^6\,\cos^3\theta\,\frac{\partial ^2}{\partial {p_\theta}^2} \frac{\partial \rho}{\partial p_r}\\
& + k^2\,{p_\theta}^2\,{r}^4\,\cos^3\theta\,\frac{\partial ^2}{\partial {p_\theta}^2} \frac{\partial \rho}{\partial p_r} -k^2\,{p_\phi}^2\,{r}^4\,\cos \theta\,\frac{\partial ^2}{\partial {p_\theta}^2} \frac{\partial \rho}{\partial p_r} +k\,{p_r}^2\,{r}^4\,\cos^3\theta\,\frac{\partial ^2}{\partial {p_\theta}^2} \frac{\partial \rho}{\partial p_r} -k\,{p_\theta}^2\,{r}^2\,\cos^3\theta\,\frac{\partial ^2}{\partial {p_\theta}^2} \frac{\partial \rho}{\partial p_r}\\
& + k\,{p_\phi}^2\,{r}^2\,\cos \theta\,\frac{\partial ^2}{\partial {p_\theta}^2} \frac{\partial \rho}{\partial p_r} -2\,\sin \theta\,k^2\,p_r\,p_\phi\,{r}^5\,\cos^2\theta\,\frac{\partial }{\partial p_\phi} \frac{\partial }{\partial p_\theta} \frac{\partial \rho}{\partial p_r} +4\,k^2\,p_\theta\,p_\phi\,{r}^4\,\cos^3\theta\,\frac{\partial }{\partial p_\phi} \frac{\partial }{\partial p_\theta} \frac{\partial \rho}{\partial p_r} \\
& + 2\,\sin \theta\,k\,p_r\,p_\phi\,{r}^3\,\cos^2\theta\,\frac{\partial }{\partial p_\phi} \frac{\partial }{\partial p_\theta} \frac{\partial \rho}{\partial p_r} -4\,k\,p_\theta\,p_\phi\,{r}^2\,\cos^3\theta\,\frac{\partial }{\partial p_\phi} \frac{\partial }{\partial p_\theta} \frac{\partial \rho}{\partial p_r} -k^2\,{p_r}^2\,{r}^6\,{\cos \theta}^5\,\frac{\partial ^2}{\partial {p_\phi}^2} \frac{\partial \rho}{\partial p_r} \\
& + \sin \theta\,k^2\,p_r\,p_\theta\,{r}^5\,{\cos \theta}^4\,\frac{\partial ^2}{\partial {p_\phi}^2} \frac{\partial \rho}{\partial p_r} -k^2\,{p_\theta}^2\,{r}^4\,{\cos \theta}^5\,\frac{\partial ^2}{\partial {p_\phi}^2} \frac{\partial \rho}{\partial p_r} +k^2\,{p_\phi}^2\,{r}^4\,\cos^3\theta\,\frac{\partial ^2}{\partial {p_\phi}^2} \frac{\partial \rho}{\partial p_r} \\
& + k\,{p_r}^2\,{r}^4\,{\cos \theta}^5\,\frac{\partial ^2}{\partial {p_\phi}^2} \frac{\partial \rho}{\partial p_r} -\sin \theta\,k\,p_r\,p_\theta\,{r}^3\,{\cos \theta}^4\,\frac{\partial ^2}{\partial {p_\phi}^2} \frac{\partial \rho}{\partial p_r} +k\,{p_\theta}^2\,{r}^2\,{\cos \theta}^5\,\frac{\partial ^2}{\partial {p_\phi}^2} \frac{\partial \rho}{\partial p_r} \\
& - k\,{p_\phi}^2\,{r}^2\,\cos^3\theta\,\frac{\partial ^2}{\partial {p_\phi}^2} \frac{\partial \rho}{\partial p_r} +k^3\,p_r\,p_\theta\,{r}^8\,\cos^3\theta\,\frac{\partial  \rho^3}{\partial {p_\theta}^3}-2\,k^2\,p_r\,p_\theta\,{r}^6\,\cos^3\theta\,\frac{\partial  \rho^3}{\partial {p_\theta}^3}+\sin \theta\,k^2\,{p_\phi}^2\,{r}^5\,\frac{\partial  \rho^3}{\partial {p_\theta}^3}\\
& + k\,p_r\,p_\theta\,{r}^4\,\cos^3\theta\,\frac{\partial  \rho^3}{\partial {p_\theta}^3}-\sin \theta\,k\,{p_\phi}^2\,{r}^3\,\frac{\partial  \rho^3}{\partial {p_\theta}^3}+k^3\,p_r\,p_\phi\,{r}^8\,\cos^3\theta\,\frac{\partial }{\partial p_\phi} \frac{\partial  \rho^2}{\partial {p_\theta}^2} -2\,k^2\,p_r\,p_\phi\,{r}^6\,\cos^3\theta\,\frac{\partial }{\partial p_\phi} \frac{\partial  \rho^2}{\partial {p_\theta}^2}\\
& - 2\,\sin \theta\,k^2\,p_\theta\,p_\phi\,{r}^5\,\cos^2\theta\,\frac{\partial }{\partial p_\phi} \frac{\partial  \rho^2}{\partial {p_\theta}^2}+k\,p_r\,p_\phi\,{r}^4\,\cos^3\theta\,\frac{\partial }{\partial p_\phi} \frac{\partial  \rho^2}{\partial {p_\theta}^2} +2\,\sin \theta\,k\,p_\theta\,p_\phi\,{r}^3\,\cos^2\theta\,\frac{\partial }{\partial p_\phi} \frac{\partial  \rho^2}{\partial {p_\theta}^2}\\
& + k^3\,p_r\,p_\theta\,{r}^8\,{\cos \theta}^5\,\frac{\partial ^2}{\partial {p_\phi}^2} \frac{\partial \rho}{\partial p_\theta} -2\,k^2\,p_r\,p_\theta\,{r}^6\,{\cos \theta}^5\,\frac{\partial ^2}{\partial {p_\phi}^2} \frac{\partial \rho}{\partial p_\theta}+\sin \theta\,k^2\,{p_\theta}^2\,{r}^5\,{\cos \theta}^4\,\frac{\partial ^2}{\partial {p_\phi}^2} \frac{\partial \rho}{\partial p_\theta} \\
& - \sin \theta\,k^2\,{p_\phi}^2\,{r}^5\,\cos^2\theta\,\frac{\partial ^2}{\partial {p_\phi}^2} \frac{\partial \rho}{\partial p_\theta} +k\,p_r\,p_\theta\,{r}^4\,{\cos \theta}^5\,\frac{\partial ^2}{\partial {p_\phi}^2} \frac{\partial \rho}{\partial p_\theta} -\sin \theta\,k\,{p_\theta}^2\,{r}^3\,{\cos \theta}^4\,\frac{\partial ^2}{\partial {p_\phi}^2} \frac{\partial \rho}{\partial p_\theta} \\
& + \sin \theta\,k\,{p_\phi}^2\,{r}^3\,\cos^2\theta\,\frac{\partial ^2}{\partial {p_\phi}^2} \frac{\partial \rho}{\partial p_\theta} +k^3\,p_r\,p_\phi\,{r}^8\,{\cos \theta}^5\,\frac{\partial  \rho^3}{\partial {p_\phi}^3} -2\,k^2\,p_r\,p_\phi\,{r}^6\,{\cos \theta}^5\,\frac{\partial  \rho^3}{\partial {p_\phi}^3}\\
& \left. +\sin \theta\,k^2\,p_\theta\,p_\phi\,{r}^5\,{\cos \theta}^4\,\frac{\partial  \rho^3}{\partial {p_\phi}^3}+k\,p_r\,p_\phi\,{r}^4\,{\cos \theta}^5\,\frac{\partial  \rho^3}{\partial {p_\phi}^3} -\sin \theta\,k\,p_\theta\,p_\phi\,{r}^3\,{\cos \theta}^4\,\frac{\partial  \rho^3}{\partial {p_\phi}^3} \right).
\end{split}
\end{equation}
If we retain just the terms linear in $k$, the $\mathcal{O}(\hbar^2)$ correction to the equation is approximated in the low curvature regime by

\begin{equation}
\begin{split}
&4\hbar^2k\left( \frac{\,{p_\theta}^2\,}{3\,{r}^3} \frac{\partial \rho^3}{\partial {p_r}^3}+\,{p_\phi}^2\,\cos \theta\,\frac{\partial  \rho^3}{\partial {p_r}^3}-2\,\,p_r\,p_\theta\,{r}^2\,\cos^3\theta\,\frac{\partial }{\partial p_\theta} \frac{\partial  \rho^2}{\partial {p_r}^2} -\sin \theta\,\,{p_\phi}^2\,r\,\frac{\partial }{\partial p_\theta} \frac{\partial  \rho^2}{\partial {p_r}^2}\right.\\
& - 2\,\,p_r\,p_\phi\,{r}^2\,\cos^3\theta\,\frac{\partial }{\partial p_\phi} \frac{\partial  \rho^2}{\partial {p_r}^2} +\,{p_r}^2\,{r}^4\,\cos^3\theta\,\frac{\partial ^2}{\partial {p_\theta}^2} \frac{\partial \rho}{\partial p_r} -\,{p_\theta}^2\,{r}^2\,\cos^3\theta\,\frac{\partial ^2}{\partial {p_\theta}^2} \frac{\partial \rho}{\partial p_r}
+\,{p_\phi}^2\,{r}^2\,\cos \theta\,\frac{\partial ^2}{\partial {p_\theta}^2} \frac{\partial \rho}{\partial p_r} \\
& + 2\,\sin \theta\,\,p_r\,p_\phi\,{r}^3\,\cos^2\theta\,\frac{\partial }{\partial p_\phi} \frac{\partial }{\partial p_\theta} \frac{\partial \rho}{\partial p_r} -4\,\,p_\theta\,p_\phi\,{r}^2\,\cos^3\theta\,\frac{\partial }{\partial p_\phi} \frac{\partial }{\partial p_\theta} \frac{\partial \rho}{\partial p_r} +\,{p_r}^2\,{r}^4\,{\cos \theta}^5\,\frac{\partial ^2}{\partial {p_\phi}^2} \frac{\partial \rho}{\partial p_r} \\
& - \sin \theta\,\,p_r\,p_\theta\,{r}^3\,{\cos \theta}^4\,\frac{\partial ^2}{\partial {p_\phi}^2} \frac{\partial \rho}{\partial p_r} +\,{p_\theta}^2\,{r}^2\,{\cos \theta}^5\,\frac{\partial ^2}{\partial {p_\phi}^2} \frac{\partial \rho}{\partial p_r} -\,{p_\phi}^2\,{r}^2\,\cos^3\theta\,\frac{\partial ^2}{\partial {p_\phi}^2} \frac{\partial \rho}{\partial p_r} +\,p_r\,p_\theta\,{r}^4\,\cos^3\theta\,\frac{\partial  \rho^3}{\partial {p_\theta}^3} \\
& - \sin \theta\,\,{p_\phi}^2\,{r}^3\,\frac{\partial  \rho^3}{\partial {p_\theta}^3} +\,p_r\,p_\phi\,{r}^4\,\cos^3\theta\,\frac{\partial }{\partial p_\phi} \frac{\partial  \rho^2}{\partial {p_\theta}^2} +2\,\sin \theta\,\,p_\theta\,p_\phi\,{r}^3\,\cos^2\theta\,\frac{\partial }{\partial p_\phi} \frac{\partial  \rho^2}{\partial {p_\theta}^2} +\,p_r\,p_\theta\,{r}^4\,{\cos \theta}^5\,\frac{\partial ^2}{\partial {p_\phi}^2} \frac{\partial \rho}{\partial p_\theta} \\
& \left.-\sin \theta\,\,{p_\theta}^2\,{r}^3\,{\cos \theta}^4\,\frac{\partial ^2}{\partial {p_\phi}^2} \frac{\partial \rho}{\partial p_\theta} +\sin \theta\,\,{p_\phi}^2\,{r}^3\,\cos^2\theta\,\frac{\partial ^2}{\partial {p_\phi}^2} \frac{\partial \rho}{\partial p_\theta}+\,p_r\,p_\phi\,{r}^4\,{\cos \theta}^5\,\frac{\partial  \rho^3}{\partial {p_\phi}^3} -\sin \theta\,\,p_\theta\,p_\phi\,{r}^3\,{\cos \theta}^4\,\frac{\partial  \rho^3}{\partial {p_\phi}^3}\right).
\end{split}
\end{equation}

\twocolumn

%

%

\bibliographystyle{IEEEtran} 
\bibliography{bibliografia}

\end{document}